\documentclass[%
 reprint,
 superscriptaddress,
 nofootinbib,
 amsmath,amssymb,
 aps,
 prd,
]{revtex4-2}

\usepackage[utf8]{inputenc}
\usepackage[english]{babel}
\usepackage{amsmath,latexsym}
\usepackage{csquotes}
\usepackage{graphicx}
\usepackage{xcolor}
\usepackage[colorlinks=true,linkcolor=magenta,citecolor=blue]{hyperref}
\usepackage{subcaption}
\usepackage[dvipsnames]{xcolor}
\usepackage{ulem}

\DeclareCaptionJustification{fulljust}{\leftskip=0pt\rightskip=0pt\parfillskip=0pt plus 1fil\relax}
\newcommand{\dd}{\mathrm{d}}
\newcommand{\fmode}[1]{{}^{#1}\!f}

\begin{document}

\title{Viscous damping of axisymmetric oscillation modes of uniformly rotating relativistic stars}

\author{Óscar H. Petit}
\affiliation{Departamento de Astronom\'{\i}a y Astrof\'{\i}sica, Universitat de Val\`encia, Avinguda Vicent
Andrés Estellés 19, 46100, Burjassot (Val\`encia), Spain.}
\email{oscar.hervas@uv.es}

\author{José A. Font}
\affiliation{Departamento de Astronom\'{\i}a y Astrof\'{\i}sica, Universitat de Val\`encia, Avinguda Vicent
Andrés Estellés 19, 46100, Burjassot (Val\`encia), Spain.}
\affiliation{Observatori Astron\`omic, Universitat de Val\`encia, C/ Catedr\'atico Jos\'e Beltr\'an 2, 46980, Paterna (Val\`encia), Spain.}


\date{\today}

\begin{abstract}
We study the effects of bulk and shear viscosity on the damping of axisymmetric oscillations of uniformly rotating relativistic stars, using the first-order viscous hydrodynamics formulation of Bemfica, Disconzi, Noronha and Kovtun (BDNK). The equations of motion are implemented in a two-dimensional numerical code under the Cowling approximation, assuming the fluid evolves on the fixed spacetime of a rotating star. We build two sequences of polytropic stellar models and investigate how damping depends on rotation rate (from the non-rotating case to the mass-shedding limit) and compactness. Our study focuses on the quasi-radial $l=0$ $F$ mode and the $l=1,2,3$ fundamental modes, covering a dataset of 288 perturbed stellar models. For all modes and chosen viscosities, the BDNK runs show that oscillations decay within a few milliseconds, much faster than in the perfect fluid case. The $F$ mode damping is dominated by bulk viscosity, consistent with its global compression of the star, while the $l=2$ and $l=3$ modes are mainly damped by shear viscosity. The $l=1$ mode is largely insensitive to both viscosity coefficients, and its decay rate is dominated by numerical dissipation. Rotation renders the $F$ mode sensitive to shear viscosity. Since the inherent numerical damping does not affect the slopes of the decay rates, our viscous sensitivity estimates faithfully identify the dominant damping mechanism for each mode, shear or bulk viscosity, and reveal robust trends with rotation rate and compactness.
\end{abstract}

\maketitle

\section{Introduction}
\label{sec:intro}

The oscillation frequencies of neutron stars encode information about the equation
of state (EoS) of dense matter, making oscillations excited in violent astrophysical
events a powerful observational probe; this is the idea behind gravitational-wave (GW)
asteroseismology~\cite{Andersson:1998,Kokkotas:1999}. Such oscillations can be
triggered by core collapse supernovae~\cite{Mezzacappa2024, Muller2026}, magnetar 
giant flares~\cite{Sotani2024}, phase transitions to quark
matter~\cite{Orsaria2019, Li2024}, and binary neutron star 
mergers~\cite{Ecker2025}. Since GW170817~\cite{LIGO:2017_GW170817}, much
attention has turned to merger remnants, hot and rapidly, differentially rotating
objects whose dominant oscillation frequencies correlate with the stellar radius and
compactness~\cite{Bauswein:2012, Takami:2014, Baiotti:2017_NSreview}, and whose damping
timescales are governed by transport properties such as shear and bulk 
viscosity~\cite{Cutler:1987, CutlerLindblomSplinter:1990, Sawyer:1989, Alford:2018}. Viscous
effects may also imprint on the GW signal during the inspiral~\cite{Ripley:2023,
Hegade:2024}, and GW170817 already places constraints on the viscosity of the
inspiralling stars~\cite{Ripley:2024}. Detecting such pulsations will thus yield
invaluable information about the internal structure and transport properties of
relativistic stars (see~\cite{Andersson2021, Kruger2021} for recent reviews),
with dramatically improved prospects from third-generation detectors such as
Cosmic Explorer and the Einstein Telescope \citep{Walker2024}.

Numerical simulations can provide detailed insights into the dynamics of neutron star oscillations.
However, including viscosity in relativistic simulations is not straightforward.
The first-order theories of Eckart~\cite{Eckart:1940} and of Landau and
Lifshitz~\cite{Landau:1959} are acausal, and their equilibrium states are
unstable~\cite{Hiscock:1985}, so they cannot be evolved numerically. The
usual way around this has been to treat the dissipative fluxes as new
fields that relax towards their Navier--Stokes values, as in the second-order theory of M\"uller, Israel and Stewart
(MIS)~\cite{Muller:1967,Israel:1976,Israel:1979}. Neutron-star
simulations have mostly relied on formulations of this kind or on
effective prescriptions. They have been used to model the transport of
angular momentum by magnetohydrodynamic turbulence in merger
remnants~\cite{Duez:2004,Shibata:2017_differential,Shibata:2017_viscous,Radice:2017_LES,Duez:2020}
and, more recently, the bulk viscosity caused by weak interactions out of
equilibrium, in spherically symmetric
evolutions~\cite{Camelio:2023a,Camelio:2023b} and in merger
simulations~\cite{Most:2022_bulk,ChabanovRezzolla:2025_PRD,ChabanovRezzolla:2025_PRL},
using MIS-type formulations~\cite{Chabanov:2021,Bemfica:2019_IS} or
evolving the reacting particle species directly. Estimates based on the
microphysics of dense
matter indicate that this bulk viscosity may be large enough to damp density
oscillations in merger remnants~\cite{Alford:2018,Most:2021}, whereas the
shear viscosity of cold neutron-star
matter~\cite{ShterninYakovlev:2008,ManuelTolos:2013} is much smaller.

In a series of papers, Bemfica, Disconzi and Noronha~\cite{Bemfica:2018,Bemfica:2019,Bemfica:2022}
and Kovtun~\cite{Kovtun:2019,Hoult:2020} (BDNK hereafter) showed that the problems of
first-order theories are tied to the choice of hydrodynamic frame. With a
more general definition of the temperature and velocity of a fluid out of
equilibrium, the first-order equations become causal, strongly hyperbolic
and stable, without additional fields ~\cite{Disconzi:2024}. Since these seminal works, the BDNK equations have
been evolved numerically in flat spacetime, for shocks and ideal-gas
microphysics~\cite{Pandya:2021,Pandya:2022,Pandya:2022b}, for conformal
fluids in comparison with holography and other
formulations~\cite{Bantilan:2022,BeaFigueras:2024, KeeblePretorius:2025}; a related dissipative extension
of ideal hydrodynamics was proposed by~\cite{HattonHawke:2024}.
Moreover, the first nonlinear evolutions of spherically symmetric neutron stars with the BDNK formulation
have been recently reported by~\cite{Shum:2025}, who studied purely radial
oscillations in the Cowling
approximation. Even more recently, a new study has presented flux-conservative 3+1 formulations of the BDNK equations
in arbitrary spacetimes~\cite{Patel:2026}.

In addition, the linearized theory of perturbations of spherically symmetric self-gravitating fluids in the BDNK approach has been reported in~\cite{RedondoYuste:2025} and the linear (radial) stability
of spherical stars within the Eckart, MIS, and BDNK theories has been 
established in~\cite{CaballeroYunes:2025}. Furthermore, the radial modes of
viscous stars have been computed in the Navier--Stokes and Israel--Stewart
theories in~\cite{Guerrieri:2026, Mendes:2026}, and in the Eckart and BDNK
theories in~\cite{KeebleRedondoYuste:2026}. The latter study also shows that
viscosity damps the radial modes on millisecond timescales and shifts
their frequencies by amounts that grow with the compactness and the
viscosity of the star. On the other hand, axial and non-radial
modes have also been studied in the linear regime by~\cite{Bussieres:2026} and~\cite{Katagiri:2026}, respectively. Moreover, 
the conditions under which a first-order description of viscous stellar oscillations is
adequate have been examined in~\cite{KeebleCardenas:2026,DenicolBreakdown:2026}. 

The aforementioned studies with the BDNK approach are restricted to either nonlinear evolutions of non-rotating stars or to linear
perturbations of spherical or slowly rotating stars. How viscosity damps the oscillation modes of rapidly
rotating stars in nonlinear evolutions of the BDNK equations has not yet been studied. This paper is the first attempt in that direction. To do so, as~\cite{Shum:2025} did in the spherically symmetric case, we adopt the simplifying assumption of the Cowling approximation~\cite{Cowling:1941}, in which the spacetime fields are
kept fixed and only the fluid evolves, albeit in a fully nonlinear fashion. Our work builds on the study of Font et al.~\cite{Font:2001_Cowling} in the inviscid case. Extending earlier
work by~\cite{Font:2000_rotNS}, \cite{Font:2001_Cowling} computed the frequencies of the quasi-radial and non-radial
axisymmetric modes of a sequence of uniformly rotating polytropes with
polytropic index $N=1$, from the non-rotating star to close to the
mass-shedding limit. Those results have been used to test many
relativistic hydrodynamics
codes~\cite{Font:2002_longterm,Stergioulas:2004_nonlinear,2006MNRAS.368.1609D,Gaertig:2008}.
For us they serve two purposes: they test the perfect-fluid limit of our
code, and they provide a set of well-identified modes whose damping we
can analyse.

In this paper we present a two-dimensional code that evolves the BDNK equations
for axisymmetric flows on the spacetime of a uniformly rotating star, and use it to study how
shear and bulk viscosity damp the quasi-radial $l=0$ mode and the $l=1,2,3$ fundamental
modes along the sequence of Ref.~\cite{Font:2001_Cowling}. Our aim is qualitative:
we seek to disentangle the respective roles of shear and bulk viscosity in damping
each mode, and to understand how these change with rotation and stellar compactness,
rather than to provide damping rates for realistic neutron stars. To this end, we
evolve each model with shear and bulk viscosities separately, and characterise each
mode by the sensitivity of its decay rate to each viscosity coefficient. This
approach isolates the viscous contribution to the damping from other sources,
including numerical dissipation, so that the results serve to establish trends rather
than to predict damping rates for real neutron stars. We also examine how these
trends change along a second sequence of increasing compactness at a
fixed polar-to-equatorial axis ratio, $r_p/r_e=0.85$.

This paper is organized as follows: Section~\ref{sec:methods} describes the formulation and the numerical code, and Section~\ref{sec:setup} the stellar models and the analysis of the simulations. We present our results in Section~\ref{sec:results} and our conclusions in Section~\ref{sec:discussion}. A resolution study of our runs is presented in Appendix~\ref{app:convergence}. We use geometrized units with $c=G=M_\odot=1$ unless stated otherwise.

\section{Formulation and numerical methods}
\label{sec:methods}

\subsection{Spacetime and stress--energy tensor}
\label{sec:formalism}

We evolve the fluid on the stationary, axisymmetric spacetime of a
uniformly rotating star, which is kept fixed during the evolution. In
the quasi-isotropic coordinates of the equilibrium models
(Sec.~\ref{sec:models}), written in cylindrical coordinates
$(\varpi,z,\phi)$, the line element reads
\begin{equation}
\dd s^2 = -\alpha^2\dd t^2 + \gamma_{ij}(\dd x^i+\beta^i\dd t)(\dd x^j+\beta^j\dd t),
\end{equation}
where $\alpha$ is the lapse function, the shift vector $\beta^i$ has only an azimuthal component,
$\beta^\phi=-\omega$, given by the frame-dragging angular velocity, and the
spatial metric $\gamma_{ij}$ is diagonal with $\gamma_{\varpi\varpi}=\gamma_{zz}$. The
Cowling approximation removes the coupling of the fluid to metric
perturbations. It shifts the frequencies of the quasi-radial modes, but
leaves the mode structure intact and allows a direct comparison with
Ref.~\cite{Font:2001_Cowling}.

For a fluid without a conserved particle current, as appropriate for the
barotropic EoS used here, the BDNK stress--energy
tensor~\cite{Bemfica:2018,Bemfica:2019,Bemfica:2022,Kovtun:2019,Hoult:2020}
contains all terms up to first order in derivatives,
\begin{align}
T^{\mu\nu} =\; & (e+\mathcal{A})\,u^\mu u^\nu + (P+\Pi)\,\Delta^{\mu\nu}
 + \mathcal{Q}^\mu u^\nu + \mathcal{Q}^\nu u^\mu \nonumber\\
 & - 2\eta\,\sigma^{\mu\nu},
\label{eq:Tmunu}
\end{align}
where $e$ is the energy density, $P$ the pressure, $u^\mu$ the
four-velocity, $\Delta^{\mu\nu}=g^{\mu\nu}+u^\mu u^\nu$ the projector
orthogonal to $u^\mu$ and $\sigma^{\mu\nu}$ the shear tensor. The
first-order corrections are
\begin{align}
\mathcal{A} &= \tau_e\left[u^\alpha\nabla_\alpha e + (e+P)\nabla_\alpha u^\alpha\right],\\
\Pi &= -\zeta\,\nabla_\alpha u^\alpha + \tau_P\left[u^\alpha\nabla_\alpha e + (e+P)\nabla_\alpha u^\alpha\right],\\
\mathcal{Q}^\mu &= \tau_Q\left[(e+P)\,u^\alpha\nabla_\alpha u^\mu + \Delta^{\mu\alpha}\nabla_\alpha P\right],
\end{align}
with the shear and bulk viscosity coefficients denoted by $\eta$ and $\zeta$. The terms in square
brackets vanish for solutions of the perfect-fluid equations and do not
change the physics at first order; they encode the choice of
hydrodynamic frame. For suitable values of the frame parameters $\tau_e$,
$\tau_P$ and $\tau_Q$, the equations $\nabla_\mu T^{\mu\nu}=0$ are causal,
strongly hyperbolic and stable around
equilibrium~\cite{Bemfica:2019,Bemfica:2022}.

\subsection{Viscosity, frame and equation of state}
\label{sec:viscosity}

We use the parametrisation and frame of Ref.~\cite{Shum:2025}. The
viscosities are proportional to the enthalpy density and to the square of
the sound speed,
\begin{equation}
\eta=\hat q\,\eta_0\,(e+P)\,c_s^2,\qquad \zeta=\hat q\,\zeta_0\,(e+P)\,c_s^2,
\label{eq:visc}
\end{equation}
and the frame parameters are
\begin{equation}
\tau_e=\tfrac{4}{3}\eta_0+\zeta_0,\qquad \tau_P=\hat sc_s^2\,\tau_e,\qquad \tau_Q=\hat a\tau_e,
\label{eq:frame}
\end{equation}
with constant $\eta_0$ and $\zeta_0$ and $\hat q=0.999$,
we use the dimensionless variables $\hat a=\hat s=1$.
The characteristic speeds are then $c_+=\sqrt{3}\,c_s$, $c_-=0.0183\,c_s$
and $c_0\simeq0.9995\,c_s\sqrt{\eta_0/(\tfrac{4}{3}\eta_0+\zeta_0)}$. The BDNK 
system is strongly hyperbolic for $\hat q<1$ and $\eta_0>0$, and causal
for $c_s^2<1/3$. Therefore, the shear viscosity cannot be entirely switched off,
which is why our bulk-viscous evolutions also contain a small amount of shear
viscosity (Sec.~\ref{sec:models}).

Both viscosities are
largest in the core of the neutron star and vanish at the surface. At the centre of the
stellar models of our rotation sequence, the value $\eta_0=0.01$ corresponds in physical units to
$\eta_c\simeq8.95\times10^{27}$\,g\,cm$^{-1}$\,s$^{-1}$ , and the largest value
used, $\eta_0=0.08$, to $\eta_c\simeq7.2\times10^{28}$\,g\,cm$^{-1}$\,s$^{-1}$.

Similar correspondences apply to the bulk viscosity coefficient. These values lie far above microphysical estimates of the shear
viscosity. In cold neutron-star cores, the scattering of electrons, muons
and neutrons gives $\eta\sim10^{16}$--$10^{20}$\,g\,cm$^{-1}$\,s$^{-1}$ at
nuclear density for temperatures of
$10^{10}$--$10^{8}$\,K~\cite{Cutler:1987,ShterninYakovlev:2008}. Even in hot
binary neutron star merger remnants, where trapped neutrinos increase it, shear viscosity damps
flows over kilometre scales only on timescales of tens of
seconds~\cite{Alford:2018}, and the averaged shear viscosity of the stars estimated in
GW170817 is bounded by
$\eta\lesssim10^{28}$\,g\,cm$^{-1}$\,s$^{-1}$~\cite{Ripley:2024}. For the
bulk viscosity, our values lie at the lower end of the range
$10^{28}$--$10^{30}$\,g\,cm$^{-1}$\,s$^{-1}$ estimated for weak interactions
in merger remnants (see Ref.~\cite{KeebleCardenas:2026} and references
therein), below the GW170817 bound
$\zeta\lesssim10^{31}$\,g\,cm$^{-1}$\,s$^{-1}$~\cite{Ripley:2024}. Such large
viscosities are needed for the viscous damping to be measurable within our
evolution times. We note, however, that since the damping is linear in the viscosity, the damping
coefficients we derive in this work can be rescaled to other values.

Following~\cite{Font:2001_Cowling} the EoS describing our stellar models is a polytrope, $P=K\rho^{1+1/N}$, with polytropic index $N=1$ and $K=100$, $\rho$ being the rest-mass density. Since
the BDNK system evolves only energy and momentum \cite{Shum:2025}, the pressure has to be
expressed in terms of the energy density. With $e=\rho+K\rho^2$,
\begin{align}
P(e)&=\frac{1+2Ke-\sqrt{1+4Ke}}{2K},\nonumber\\
c_s^2&=\frac{\dd P}{\dd e}=1-\frac{1}{\sqrt{1+4Ke}},
\label{eq:eos}
\end{align}
which is the same $N=1$ polytrope written with respect to a different variable.

\subsection{Evolution equations}

The simulations discussed in this paper are performed with a BDNK code that we developed from scratch for this work. The equations
$\nabla_\mu T^{\mu\nu}=0$ are cast in flux-conservative 3+1
form~\cite{Valencia,Font:2008_review}
\begin{equation}
\partial_t\left(\sqrt{\gamma}\,\mathbf{U}\right)
+\partial_i\left(\sqrt{\gamma}\,\mathbf{F}^i\right)=\sqrt{\gamma}\,\mathbf{S},
\label{eq:conservation}
\end{equation}
where $\gamma$ is the determinant of the spatial metric $\gamma_{ij}$ and
$\mathbf{U}=(E,S_\varpi,S_z,S_\phi)$ are the conserved variables. They are defined as 
$E=n_\mu n_\nu T^{\mu\nu}$ and $S_j=-n_\mu\gamma_{\nu j}T^{\mu\nu}$ and represent the
energy and momentum densities measured by the Eulerian observer, where $n^\mu$, with $n_\mu=(-\alpha,0,0,0)$, is the unit normal to the
hypersurfaces of constant $t$. The source
terms $\mathbf{S}$ contain the derivatives of the background metric;
the interested reader is addressed to~\cite{Shum:2025} for the explicit form of the source terms.
The fluxes in Eq.~(\ref{eq:conservation}) are
\begin{equation}
\mathbf{F}^i=\left(\alpha S^i-\beta^iE,\ \alpha S^i_{\ j}-\beta^iS_j\right),
\label{eq:fluxes}
\end{equation}
with $S^i=\gamma^{ij}S_j$ and the spatial stress
$S^{ij}=\gamma^i_{\ \mu}\gamma^j_{\ \nu}T^{\mu\nu}$. 
Because the BDNK stress-energy tensor~\eqref{eq:Tmunu} contains first
derivatives of the fluid variables, including time derivatives, the
conserved variables depend not only on the primitive variables
$\mathbf{p}_0=(e,v^\varpi,v^z,v^\phi)$ but also on their time derivatives. Here
$v^i=(u^i/u^t+\beta^i)/\alpha$ is the fluid three-velocity measured by the
Eulerian observer, so that $u^\mu=W(n^\mu+v^\mu)$ with the Lorentz factor
$W=(1-\gamma_{ij}v^iv^j)^{-1/2}$. The time
derivatives are taken along $n^\mu$,
$\mathbf{p}_1=(\hat e,\hat v^\varpi,\hat v^z,\hat v^\phi)$ with
$\hat e=-n^\mu\nabla_\mu e$ and
$\hat v^i=-\gamma^i_{\ \alpha}n^\nu\nabla_\nu v^\alpha$. The code therefore
evolves the conserved variables with Eq.~\eqref{eq:conservation} and the
primitive variables with the equations that give $\partial_t\mathbf{p}_0$ in
terms of $\mathbf{p}_1$, and recovers $\mathbf{p}_1$ from
$\mathbf{U}$ after every update of $\mathbf{U}$ during the time integration.

For given $\mathbf{p}_0$ and its spatial derivatives ($\partial_\varpi e$,
$\partial_z e$, $\partial_\varpi v^i$ and $\partial_z v^i$), which are computed
from $\mathbf{p}_0$ on the grid, the conserved variables are linear in
$\mathbf{p}_1$, because the BDNK stress--energy tensor is linear in the
first derivatives. The recovery step therefore amounts to solving a
$4\times4$ linear system in each cell~\cite{Pandya:2022}, instead of the customary 
nonlinear root finding needed for a perfect fluid~\cite{Font:2008_review}.

\subsection{Discretisation}

The equations of motion are discretised on a uniform, cell-centred cylindrical
grid covering $\varpi\in[0,14]$ and $z\in[-14,14]$ in code units, with $N$
cells in each direction; we use $N=216$ for the production runs and
$N=144$, $216$ and $324$ for the resolution study.
Interface fluxes are computed with the local Lax--Friedrichs (Rusanov)
solver~\cite{Rusanov:1961}, using the largest BDNK characteristic speed in
each direction. The primitive variables are reconstructed with the
piecewise parabolic method (PPM)~\cite{Colella:1984_PPM} and the
time-derivative variables with the monotonised central (MC)
limiter~\cite{vanLeer:1977}. Time integration uses the
third-order strong-stability-preserving Runge--Kutta
scheme~\cite{Shu:1988} with a Courant number of $0.15$. Reflection symmetry is imposed across the
rotation axis. The perfect-fluid reference evolutions are computed with
the perfect-fluid version of the code on the same grid.

\subsection{Stellar surface and atmosphere}
\label{sec:surface}

The recovery of the time derivatives becomes ill conditioned where the
density and the viscosities go to zero. As the viscosities~\eqref{eq:visc}
vanish at the stellar surface, cells with rest-mass density below
$\rho_\mathrm{cut}=5\times10^{-7}$ (about $4\times10^{-4}$ of the central
density) are evolved as a perfect fluid: the viscous stress is switched
off there and the perfect-fluid recovery is used. 

Test evolutions with a different threshold show that this treatment does not affect our results
nor it is the cause of the relaxation discussed in Sec.~\ref{sec:transient}.
Outside the star the grid is filled with an atmosphere at rest with
$\rho_\mathrm{atm}=10^{-13}$, and cells whose density falls below
$100\,\rho_\mathrm{atm}$ are reset to atmosphere values. Without this
margin above the floor, near-vacuum cells with spurious velocities at the
equatorial surface of the fastest rotators made the recovery fail.

\section{Stellar models and simulations}
\label{sec:setup}
\label{sec:models}

\begin{table}[!t]
\caption{Sequence of stellar models. From left to right $r_p/r_e$ is the ratio
of polar to equatorial coordinate radii, $\Omega/\Omega_{\rm ms}$ is the
angular velocity relative to that at the mass-shedding limit of the
sequence ($r_p/r_e=0.576$, $f_{\rm rot}=854$\,Hz), $f_{\rm rot}=\Omega/2\pi$,
$M$ is the gravitational mass, $R$ is the circumferential equatorial radius, and
$j=J/M^2$ is the dimensionless angular momentum.}
\label{tab:models}
\begin{ruledtabular}
\begin{tabular}{cccccc}
$r_p/r_e$ & $\Omega/\Omega_{\rm ms}$ & $f_{\rm rot}$ [Hz] & $M\,[M_\odot]$ & $R$ [km] & $j$ \\
\hline
1.00 & 0     & 0   & 1.400 & 14.15 & 0     \\
0.95 & 0.406 & 347 & 1.432 & 14.51 & 0.222 \\
0.90 & 0.571 & 487 & 1.466 & 14.92 & 0.318 \\
0.85 & 0.692 & 591 & 1.504 & 15.38 & 0.395 \\
0.80 & 0.788 & 673 & 1.544 & 15.91 & 0.460 \\
0.75 & 0.866 & 740 & 1.586 & 16.53 & 0.518 \\
0.70 & 0.928 & 792 & 1.628 & 17.26 & 0.568 \\
0.65 & 0.972 & 830 & 1.666 & 18.16 & 0.609 \\
\end{tabular}
\end{ruledtabular}
\end{table}

\begin{table}[!t]
\caption{Compactness sequence at $r_p/r_e=0.85$,
obtained by varying the central density. $M$ is the gravitational mass,
$R$ the circumferential equatorial radius, $C=M/R$, and $c_{s,c}^2$ the
squared central sound speed.}
\label{tab:compactness}
\begin{ruledtabular}
\begin{tabular}{ccccc}
$\rho_c\,[10^{-3}]$ & $M\,[M_\odot]$ & $R$ [km] & $C$ & $c_{s,c}^2$ \\
\hline
0.60 & 1.052 & 17.54 & 0.089 & 0.107 \\
0.80 & 1.234 & 16.81 & 0.108 & 0.138 \\
1.00 & 1.369 & 16.17 & 0.125 & 0.167 \\
1.28 & 1.504 & 15.38 & 0.144 & 0.204 \\
1.60 & 1.603 & 14.62 & 0.162 & 0.242 \\
2.00 & 1.676 & 13.82 & 0.179 & 0.286 \\
\end{tabular}
\end{ruledtabular}
\end{table}

We consider two different sequences of stellar models. On the one hand, the main sequence consists of uniformly
rotating stars with central rest-mass density
$\rho_c=1.28\times10^{-3}$, from the non-rotating star to close to the mass-shedding
limit at equatorial-to-polar axis ratio $r_p/r_e=0.65$. The specific models of this sequence are reported in Table~\ref{tab:models}. On the other hand, to study the role of stellar compactness  we built a second sequence at
fixed axis ratio $r_p/r_e=0.85$ by changing the central density of the
same polytrope. Those models are described in  Table~\ref{tab:compactness}. All of the latter models lie on the
stable branch; the maximum mass, $1.73\,M_\odot$, is reached at
$\rho_c\simeq3\times10^{-3}$. Along this sequence the central sound speed
grows from $c_s^2=0.11$ to $0.286$, so the most compact model lies close to
the causality limit $c_s^2<1/3$ (Sec.~\ref{sec:viscosity}). All models were
computed with the \textsc{rns} code of Stergioulas and Friedman~\cite{Stergioulas:1995_RNS}, which solves the equilibrium equations with the method of Komatsu, Eriguchi and Hachisu~\cite{Komatsu:1989}, and interpolated onto the evolution
grid, which provides both the background spacetime and the initial fluid
state.

We excite the oscillations with the trial functions of
Ref.~\cite{Font:2001_Cowling}. For $l=0$ the rest-mass density is
perturbed by
\begin{equation}
\delta\rho=A\,\rho_c\,\sin\!\left(\frac{\pi r}{r_s(\theta)}\right),
\label{eq:pert_l0}
\end{equation}
with the energy density and pressure following from the EoS. Correspondingly, for
$l=1$, $2$ and $3$ a polar velocity is added (different in each case),
\begin{equation}
v_\theta=A\,\sin\!\left(\frac{\pi r}{r_s(\theta)}\right)\sin\theta
\times\left\{1,\;\cos\theta,\;1-5\cos^2\theta\right\},
\label{eq:pert_l123}
\end{equation}
where $r_s(\theta)$ is the coordinate radius of the stellar surface and the amplitude of the perturbation is $A=10^{-3}$. The radial factor makes the perturbations vanish at the surface.

The BDNK evolution also needs initial values of the time-derivative variables. We set the coordinate time derivatives $\partial_te$ and $\partial_tv^i$ to zero and compute the conserved variables from the complete BDNK stress--energy tensor, so that all variables are consistent with each other at $t=0$. This choice is exact for the unperturbed star but not for the perturbed one (see Sec.~\ref{sec:transient}).

Each model of the main sequence is evolved for $l=0$--$3$ in nine
configurations: the perfect fluid one, a series with only shear viscosity with coefficients
$\eta_0=0.01,0.03,0.05,0.08$ and $\zeta_0=0$, a series with mainly bulk viscosity with coefficients
$\zeta_0=0.03,0.05,0.08$ and $\eta_0=0.01$, and, finally, a run with
$\eta_0=\zeta_0=0.05$. This gives 288 evolutions of 30\,ms each. On the other hand, the models
of the compactness sequence are evolved for $l=0$--$3$ in seven configurations: the perfect fluid one,
a series with $\eta_0=0.01,0.03,0.08$ and $\zeta_0=0$, a series with $\zeta_0=0.03,0.08$ and 
$\eta_0=0.01$, and a final run with $\eta_0=\zeta_0=0.05$; the model with
$\rho_c=1.28\times10^{-3}$ is taken from the main sequence. 

The specific values of the viscosity coefficients are chosen so that the damping of the modes
can be measured within the evolution time. Their central values span the physical range 
$9\times10^{27}$--$7\times10^{28}$\,g\,cm$^{-1}$\,s$^{-1}$. Therefore, they lie below the
range $10^{30}$--$10^{31}$\,g\,cm$^{-1}$\,s$^{-1}$ at which a first-order
description of radial stellar oscillations becomes
insufficient~\cite{KeebleCardenas:2026}. They are, however, much larger than the shear viscosity of cold neutron-star
matter~\cite{Cutler:1987,ShterninYakovlev:2008}. This choice is deliberate
(Sec.~\ref{sec:viscosity}): with realistic shear viscosities, the viscous
damping would be far slower than our evolutions can follow and would be
hidden by the numerical damping. Because the viscous damping is linear in
the viscosity, our conclusions on which viscosity damps which mode, and
on how this changes with rotation and compactness, do not depend on its
magnitude.

  During the evolution, the code records the projections of the
  perturbations of the rest-mass density, the energy density and the radial
  and polar velocities onto the multipoles $l=0$--$3$. We identify the modes
  in the spectra of the perfect-fluid evolutions, near the frequencies of
  Ref.~\cite{Font:2001_Cowling}. Because rotation couples the quasi-radial and nonradial multipoles,
  each projection contains several modes; we therefore measure the
  amplitude $A(t)$ of each mode by least-squares fits of sinusoids at the
  frequencies of all identified modes, in sliding windows of 3\,ms, and
  obtain its decay rate $\Gamma$ from a weighted linear fit of $\ln A(t)$.
  The fits start after the initial transient
  (Sec.~\ref{sec:transient}), and all evolutions of a given model and mode
  are fitted over the same interval, which ends when the fastest-decaying
  evolution has dropped by a factor of 30 or approaches its noise level.

\section{Results}
\label{sec:results}

\subsection{Perfect-fluid limit}
\label{sec:validation}

\begin{figure}[!htb]
    \centering
    \includegraphics[width=\linewidth]{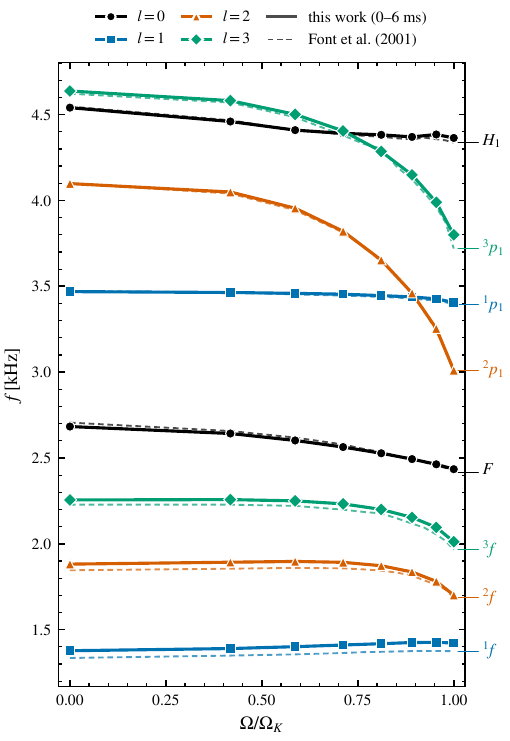}
    \caption{Frequencies of the quasi-radial ($l=0$) and $l=1,2,3$ modes
    along the rotation sequence of
    the perfect-fluid evolutions (symbols and solid lines), compared with
    those of Font et al.~\cite{Font:2001_Cowling} (dashed lines).
    Fundamental modes and first overtones are labelled on the right.}
    \label{fig:frequencies}
\end{figure}

Figure~\ref{fig:frequencies} compares the
frequencies of our perfect-fluid evolutions with those of
Ref.~\cite{Font:2001_Cowling}. The median difference is $0.6\%$ and the largest $3.7\%$. The
quasi-radial modes and all first overtones agree to better than $1\%$. The
non-radial fundamental modes are systematically higher, by $1.3\%$
($\fmode{3}$), $2.1\%$ ($\fmode{2}$) and $3.2\%$ ($\fmode{1}$), almost
independently of rotation. Our frequencies change by less than $0.1\%$ between our three
resolutions (Appendix~\ref{app:convergence}), so these offsets are not a
resolution effect of our evolutions. They are comparable to the accuracy
of $1$--$2\%$ that~\cite{Font:2001_Cowling} estimated for its
frequencies, which were obtained from 10 ms evolutions on spherical grids
of $200\times80$ and $160\times120$ points (i.e.~3 times shorter than our evolutions). The numerical schemes also
differ: Ref.~\cite{Font:2001_Cowling} evolved the rest-mass density with
Marquina's flux formula on a spherical grid, whereas we evolve the
energy density with the local Lax--Friedrichs flux on a cylindrical grid. The changes produced by rotation, including
the decrease of the $\fmode{l}$ frequencies and the avoided crossings of
the overtones near mass shedding, are reproduced along the whole sequence.

\subsection{Central density and spectra}
\label{sec:spectra}

\begin{figure}[!t]
    \centering
    \includegraphics[width=\linewidth]{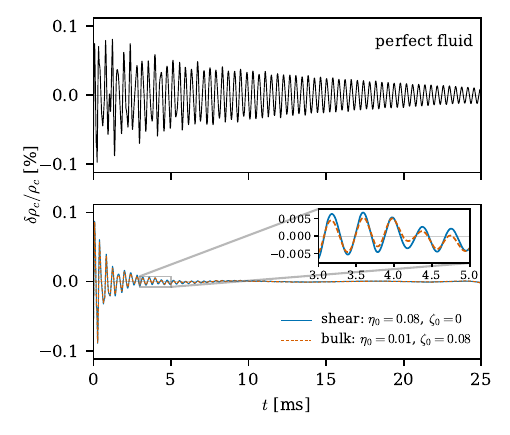}
    \caption{Relative change of the central density evolution of the 
    $r_p/r_e=0.80$ model, using the $l=0$ perturbation of Eq.~(\ref{eq:pert_l0}). The top panel corresponds to the perfect fluid and the bottom panel to 
    the strongest shear viscosity ($\eta_0=0.08$, $\zeta_0=0$, solid) and
    bulk viscosity ($\eta_0=0.01$, $\zeta_0=0.08$, dashed) cases. For the viscous cases the two lines almost overlap fully; the inset highlights the small differences. The slow drift of the central density has been removed.}
    \label{fig:timeseries}
\end{figure}

\begin{figure*}[!t]
    \centering
    \includegraphics[width=0.85\textwidth]{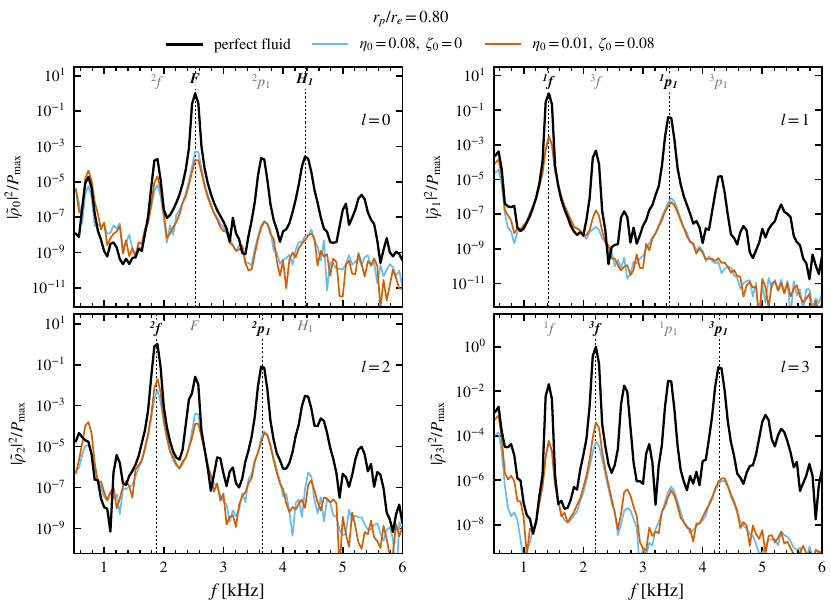}
    \caption{Power spectra of the $l=0$, 1, 2 and 3 components of the density perturbation over the whole evolution for the $r_p/r_e=0.80$ model, each from the run perturbed with the corresponding multipole: perfect fluid (black), strongest shear viscosity ($\eta_0=0.08$, $\zeta_0=0$, light blue) and strongest bulk viscosity ($\eta_0=0.01$, $\zeta_0=0.08$, orange). Bold black labels mark the modes of the $l$ multipole of each panel, grey labels modes of other multipoles that appear through rotational coupling. The power is normalised to the perfect-fluid maximum.}
    \label{fig:spectra}
\end{figure*}

Figure~\ref{fig:timeseries} shows how viscosity affects the evolutions of the $r_p/r_e=0.80$ model ($\Omega/\Omega_K=0.788$), using the $l=0$ perturbation of Eq.~(\ref{eq:pert_l0}). Both the model and the perturbation have been chosen as representative examples, as the behaviour shown in Fig.~\ref{fig:timeseries} is generic across our model sample. The figure displays the relative change of the central density evolution for both, the perfect fluid run (top panel) and the viscous runs (bottom panel). For the latter, the oscillations decay within a few milliseconds, much faster than in the perfect fluid case, where they are damped slowly by numerical viscosity; part of this faster decay is numerical (Sec.~\ref{sec:transient}), and its viscous part is what we quantify in the following sections.

Figure~\ref{fig:spectra} shows the spectra of the same 
model with $r_p/r_e=0.80$ for all $l$ perturbations, computed considering the whole evolution window. In the perfect fluid case, each projection is dominated by the fundamental mode of its multipole and by the first overtone. Because the
rotation couples different multipoles, modes of other $l$
appear too: $F$ and $H_1$ in $l=2$, $\fmode{2}$ in $l=0$, and $\fmode{1}$ and $\fmode{3}$ in each other's spectra.

Viscosity lowers all peaks by several orders of magnitude. Part of this reduction is the same for every viscous run and comes from a numerical relaxation effect (to be 
discussed in Sec.~\ref{sec:transient}). The difference between shear and bulk viscosity, however, depends on the mode. Bulk viscosity removes more power from the $F$ peak than shear viscosity of the same magnitude, while for the $\fmode{2}$ and $\fmode{3}$ modes it is the other way around: the $\fmode{3}$ peak of the bulk-viscous run is about six times higher than that of the shear-viscous run. The two viscous spectra coincide at the $\fmode{1}$ peak. The overtones, as expected, are suppressed more strongly than the fundamental modes, and the higher overtones disappear into the noise.

Viscosity, on the other hand, hardly changes the mode frequencies: the peaks of the viscous evolutions lie at the perfect-fluid frequencies, within the frequency resolution of the spectra. This agrees with linear calculations, which find frequency shifts well below a per cent for viscosities of our magnitude~\cite{KeebleRedondoYuste:2026}. The shifts reach
the per-cent level only for bulk viscosities of about
$10^{30}$\,g\,cm$^{-1}$\,s$^{-1}$~\cite{KeebleRedondoYuste:2026}. Therefore, for realistic shear viscosities, which are much smaller than ours, and for bulk viscosities below this value, the frequencies computed for perfect fluids remain accurate. We note, however, this may change in other
configurations, such as differentially rotating stars.

\subsection{Mode amplitudes}
\label{sec:amplitudes}

\begin{figure*}[!t]
    \centering
    \includegraphics[width=\textwidth]{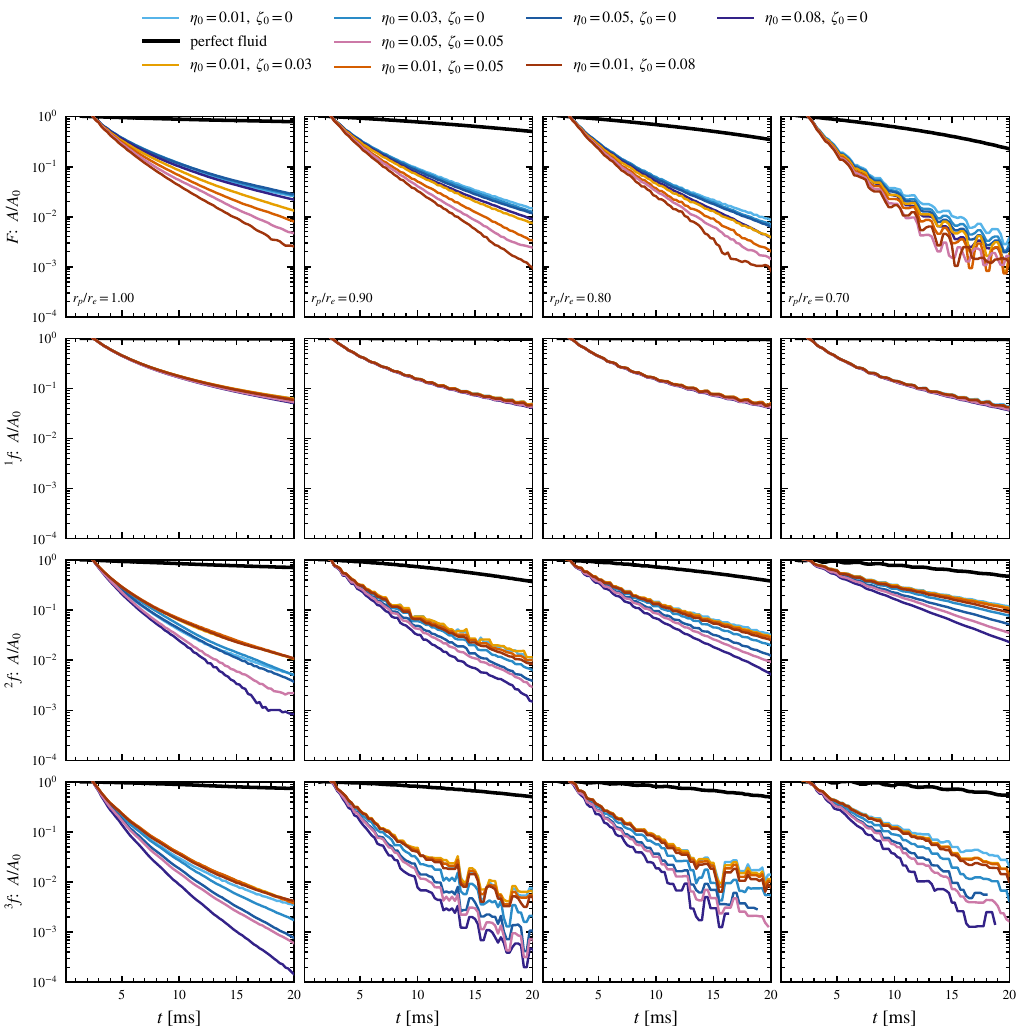}
    \caption{Amplitudes of the fundamental modes $F$, $\fmode{1}$,
    $\fmode{2}$ and $\fmode{3}$ (rows), normalised to the initial
    amplitude of the perfect-fluid run, for four rotation rates
    (columns) and all viscosities. Black: perfect fluid; blue-coloured lines: shear viscosity
    series ($\zeta_0=0$, darker for larger $\eta_0$; see also legend on top of figure); yellow to red: bulk viscosity 
    series ($\eta_0=0.01$, darker for larger $\zeta_0$); pink: combined viscosity run
    ($\eta_0=\zeta_0=0.05$).}
    \label{fig:decays}
\end{figure*}

Figure~\ref{fig:decays} shows the amplitudes of the four fundamental modes for four rotation rates $r_p/r_e=(1, 0.90, 0.80, 0.70)$. In all viscous evolutions, the amplitudes first drop quickly ($3-5$ ms) by the same amount for every viscosity, whereas the perfect-fluid amplitudes do not. This drop is a gradual decrease of the amplitude of the modes that happens at the start of the simulation and relaxes onward, rather than a sharp drop. We discuss this initial
transient in Sec.~\ref{sec:transient}. 

The amplitude of each mode decreases over time, and their ordering reveals which viscosity acts on each mode. The $F$ mode amplitudes (first row) are ordered by the bulk viscosity. In the non-rotating star, the four shear-only runs (blue-coloured lines) can hardly be told apart, while the bulk-dominated runs separate clearly, with the fastest
decay for the largest $\zeta_0$. With increasing rotation, the shear viscosity runs start to separate as well, showing that rotation enhances the activity of shear viscosity. On the other hand, for the $\fmode{2}$ and $\fmode{3}$ modes (third and fourth rows) the fastest decaying evolutions are those dominated by the shear viscosity. The evolution of the bulk-dominated runs lie on top of the $\eta_0=0.01, \zeta_0=0$ shear-only run, meaning the $l=2,3$ modes are hardly affected by bulk viscosity (note that all bulk-dominated runs have $\eta_0=0.01)$.  Finally, the behaviour of the $\fmode{1}$ amplitudes for all viscous runs
nearly coincides. This behaviour is expected since the $^1f$ mode perturbation is not compressive and does not generate shear movements; it is a movement of the whole star at once. The decay of the runs in this case is dominated by numerical damping that does not depend on the physical viscosity (Sec.~\ref{sec:transient}).

\subsection{Sensitivity of the decay rates to viscosity}
\label{sec:damping}

\begin{figure*}[!t]
    \centering
    \includegraphics[width=\textwidth]{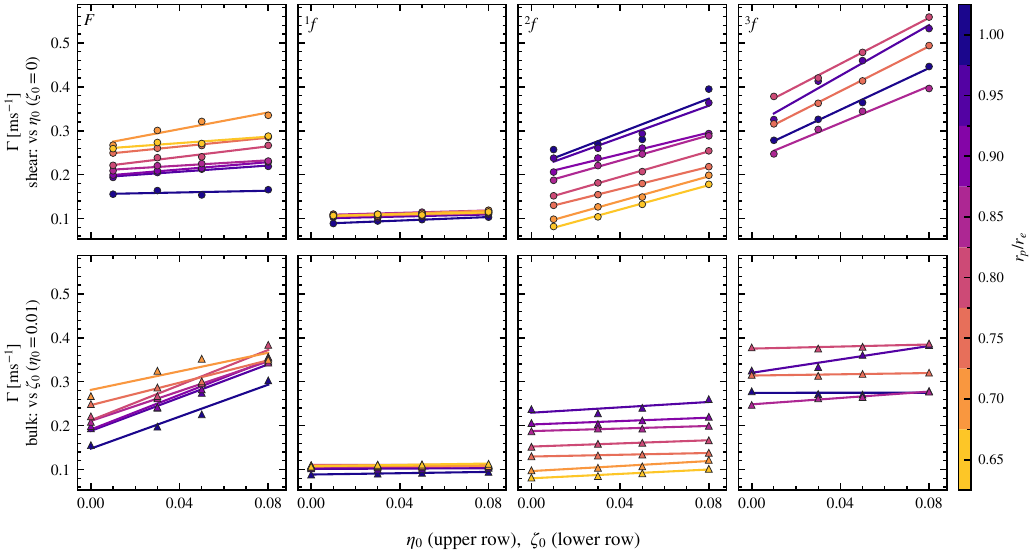}
    \caption{Decay rates of the fundamental modes as a function of the
    shear viscosity at $\zeta_0=0$ (upper row) and of the bulk viscosity
    at $\eta_0=0.01$ (lower row), for all rotation rates (colour). Lines
    are least-squares fits whose slopes are the viscous sensitivities.
    Series dominated by marginal fits are not shown.}
    \label{fig:gamma_vs_visc}
\end{figure*}

\begin{figure*}[!t]
    \centering
    \includegraphics[width=\textwidth]{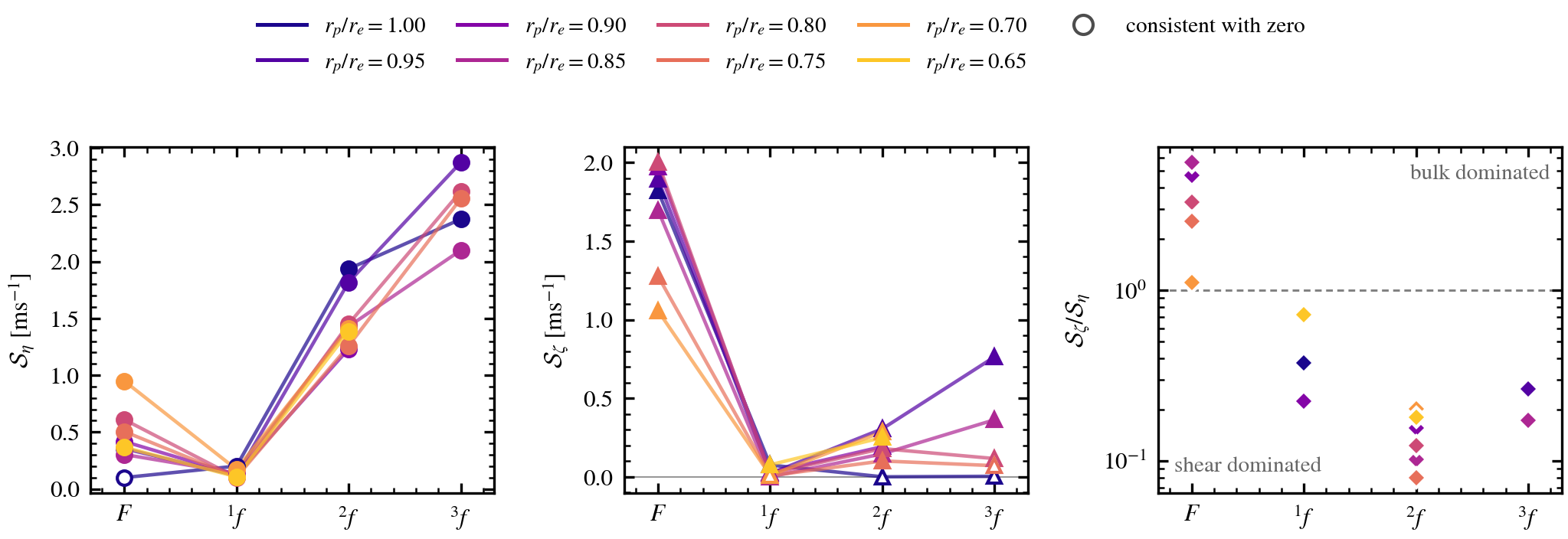}
    \caption{Shear (left) and bulk (middle) sensitivities, $\mathcal{S}_\eta$ and $\mathcal{S}_\zeta$, of the fundamental modes for all
    rotation rates (color code indicated in the figure title). Open symbols
    are consistent with zero. Their ratio is displayed on the right panel.}
    \label{fig:l_dependence}
\end{figure*}

After the initial transient, the decay rate of every mode increases approximately linearly with the viscosity coefficients, as shown in Figure~\ref{fig:gamma_vs_visc} for all stellar models of our first sequence. We describe the decay rate of each mode as
\begin{equation}
\Gamma\simeq\Gamma_{\rm 0}+\mathcal{S}_\eta\,\eta_0+\mathcal{S}_\zeta\,\zeta_0,
\ \ \ 
\mathcal{S}_\eta\equiv\frac{\partial\Gamma}{\partial\eta_0},\quad
\mathcal{S}_\zeta\equiv\frac{\partial\Gamma}{\partial\zeta_0},
\label{eq:sensitivities}
\end{equation}
where $\Gamma_{\rm 0}$ is the part of the decay rate that does not depend on viscosity. Quantities $\mathcal{S}_\eta$ and $\mathcal{S}_\zeta$ give the increase of the decay rate per unit of the shear and of the bulk viscosity coefficient, respectively. Therefore, they quantify the viscous damping. We refer to them as the shear and bulk {\it sensitivities} of the mode.

Figure~\ref{fig:l_dependence} shows both sensitivities, and their ratio, for all modes along the rotation sequence. The quasi-radial mode $F$ responds mainly to bulk viscosity. In the non-rotating star its oscillation is a pure compression: the fluid expands and contracts radially without shear, so that only bulk viscosity can damp it. We find that its bulk  sensitivity largely exceeds the shear sensitivity. When we take rotation into account this mode acquires a non-radial component, which grows with the deformation of the star and introduces shear, so that the shear sensitivity grows with the
rotation rate. Near mass shedding the motion is concentrated in the equatorial outer layers, where viscosity~\eqref{eq:visc} is small. Hence, the bulk sensitivity decreases, and the two sensitivities become
comparable.

The $\fmode{2}$ and $\fmode{3}$ modes behave in the opposite way. Their motion is nearly incompressible: the fluid moves around the star without changing its density much, so that bulk viscosity has little to act on, while the velocity gradients produce shear. The shear sensitivities of these two modes are large at all rotation rates and their bulk ones small. Finally, as mentioned before, the $\fmode{1}$ mode is barely damped by physical viscosity, making both sensitivities almost vanish for this mode.

\subsection{Numerical damping}
\label{sec:transient}

\begin{figure*}[!t]
    \centering
    \includegraphics[width=0.85\textwidth]{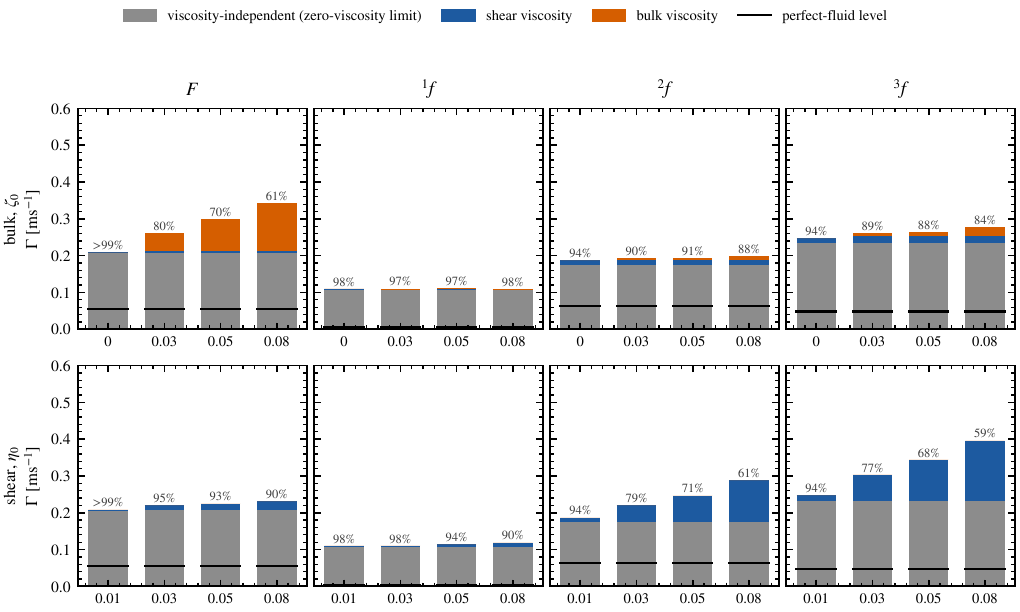}
    \caption{Decay rate budget of the fundamental oscillation modes (columns) of the $r_p/r_e=0.85$ model at $N=216$ grid resolution. The top row corresponds to the bulk-dominated series ($\eta_0=0.01$) and the bottom one to the shear-only series ($\zeta_0=0$). Grey: decay rate extrapolated to zero viscosity, which is of numerical origin. Blue and orange: damping by shear and bulk viscosity, respectively; in the bulk-dominated series, the blue part is the shear damping of $\eta_0=0.01$, which all its BDNK evolutions contain. The black horizontal lines mark the decay rate of the perfect-fluid evolution, and the percent labels give the unphysical (numerical) fraction of each rate.}
    \label{fig:budget}
\end{figure*}

Not all the measured damping is originated from physical viscosity. Numerical viscosity arising from the discretization of the evolution equations on a finite grid, hence turning the continuous PDEs into their finite difference (finite volume) version, also contributes to the damping of the stellar pulsations. Figure~\ref{fig:budget} shows each contribution to the decay rate for the  $r_p/r_e=0.85$ model, separated in its numerical part and its physical part, both shear and bulk.

To build Fig.~\ref{fig:budget} we proceed as follows. For each mode, we do a linear fit to the decay rates of the shear-viscosity runs, 
 $\Gamma=\Gamma_0+\mathcal{S}_\eta\,\eta_0$. Its constant term $\Gamma_0$ is the decay rate extrapolated to zero viscosity. This term accounts for the
damping that contributes in the absence of physical viscosity, i.e.\ the numerical part
(grey colour in Fig.~\ref{fig:budget}), and it is the same for every evolution of the same model. In the shear viscosity 
series (bottom row of the figure), the rest of each measured decay rate, $\Gamma-\Gamma_0$, is the
damping due to shear viscosity (blue colour). On the other hand, the evolutions of the bulk viscosity series, as mentioned above, also
contain some amount of shear viscosity, $\eta_0=0.01$. Hence, we attribute to this a contribution to the global damping of 
$0.01\,\mathcal{S}_\eta$
(shown in blue in the top row of Fig.~\ref{fig:budget}). The remaining contribution in this series is, thus, the
damping by bulk viscosity (orange colour in the top row).  This treatment is valid as long as we assume the numerical damping for different viscous runs of a given model is the same (as the grey-shaded areas in Fig.~\ref{fig:budget} show).

As discussed before, we observe from this figure that the bulk viscosity mainly affects the compressive $F$ mode, and its effects are highly suppressed in the higher $l=2,3$ modes. On the other hand, shear viscosity mainly acts in these high $l=2,3$ modes. Moreover, one can see in Fig.~\ref{fig:budget} that the increase of the damping rates is approximately linear with the increase of the viscous coefficients $\zeta_0, \eta_0$.

Let us now come back to Fig.~\ref{fig:decays}. All viscous evolutions begin with a numerical origin transient; during the first milliseconds the mode amplitudes drop quickly, by the same amount for every value of the viscosity coefficients, while the perfect-fluid amplitudes do not drop.
Part of it is the unavoidable dissipation of the numerical scheme, which is larger for the BDNK equations than for the perfect fluid ones. The dissipation of the Lax--Friedrichs flux is proportional to the largest characteristic speed, $\sqrt{3}\,c_s$ instead of $c_s$, and the time-derivative variables are reconstructed with a second-order accurate MC limiter. The spatial derivatives are also reconstructed, introducing even more numerical viscosity to the scheme. In the non-rotating model the central density of all BDNK evolutions decreases by about $1\%$ over 30\,ms, and by a similar amount for all values of the viscosity coefficients and rotations we consider. The density drift decreases with resolution (see Appendix~\ref{app:convergence}). We attribute it to an inconsistency between the evolved primitive and conserved variables in the outer layers of the star, where the scheme is only first order because of the treatment of the stellar surface (both the adaptive conservative-to-primitive method, which can induce small errors and the reconstruction and derivative calculation across the surface). 

\begin{figure*}[!t]
    \centering
    \includegraphics[width=\textwidth]{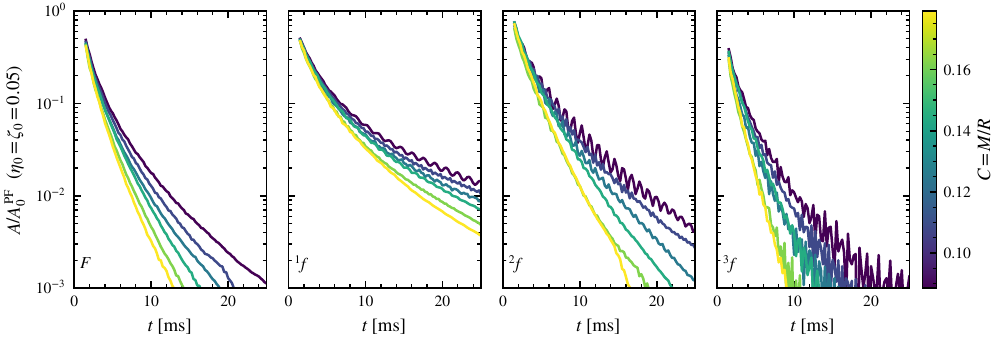}
    \caption{Amplitudes of the fundamental modes for $\eta_0=\zeta_0=0.05$
    along the compactness sequence at $r_p/r_e=0.85$
    (Table~\ref{tab:compactness}), normalised to the initial amplitude of
    the corresponding perfect-fluid run. Colours indicate the compactness
    $C=M/R$.}
    \label{fig:compactness}
\end{figure*}

We test the effects of the damping associated with the numerical viscosity in two ways. First, we extrapolate the decay rates of the non-rotating model, computed at grid resolutions of $N=144$, $216$ and $324$ over a common fit window, to infinite resolution. The perfect-fluid rates extrapolate to approximately zero, as expected for numerical dissipation. The zero-viscosity rate of the BDNK runs, however, decreases only from $0.16$
to $0.13$\,ms$^{-1}$ for the $F$ mode and from $0.25$ to $0.15$\,ms$^{-1}$ for the $\fmode{2}$ mode, so that the extrapolation does not remove the decay completely. Since the BDNK equations reduce to the perfect-fluid equations as the viscosity vanishes, this damping cannot be physical. Its origin is purely numerical and, moreover, it has a convergence rate that is slower than first order. We note that Shum et al.~\cite{Shum:2025} find a similar behaviour for the same non-rotating star studied here. Their decay rates extrapolated to infinite resolution also contain a large part that hardly depends on the physical viscosity. 

Second, we evolve the models with $r_p/r_e=1.00$, $0.85$ and $0.70$ with a smaller shear viscosity coefficient than in our fiducial shear-only models, $\eta_0=0.005$. We find that their decay rates lie on the extension of the linear fits, which implies that the unphysical damping does not depend on the shear viscosity down to this value. (Evolutions with $\eta_0\leq0.0025$  give erratic decay rates, because $\tau_e$ becomes several times shorter than the time step, and we do not use them.) This independence is restricted to the range of viscosities we can evolve: as the viscosity vanishes, the BDNK evolutions must approach the perfect-fluid ones.

\subsection{Dependence on compactness}
\label{sec:compactness}

\begin{figure*}[!t]
    \centering
    \includegraphics[width=\textwidth]{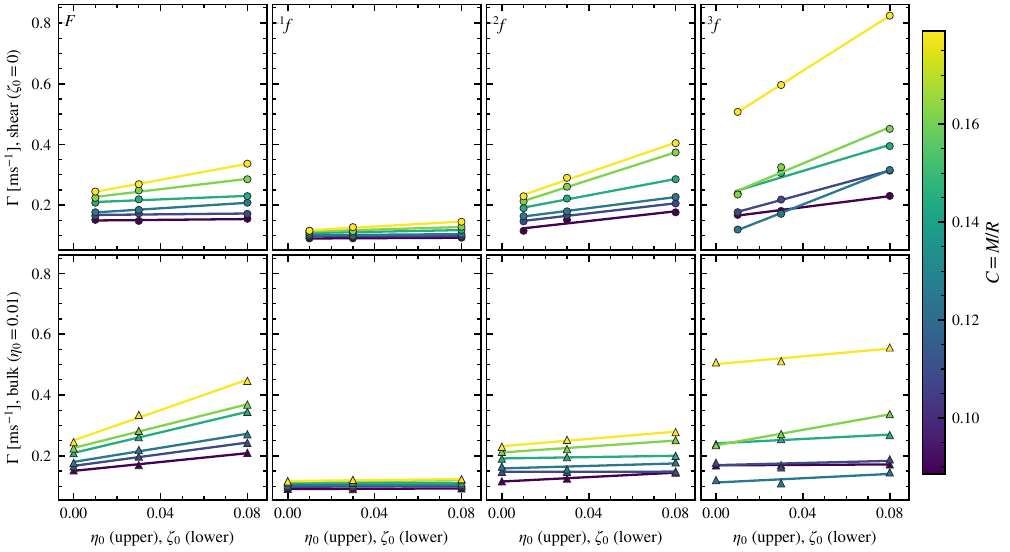}
    \caption{Decay rates of the oscillation modes along the compactness sequence (colour-coded) as a function of the shear
    viscosity at $\zeta_0=0$ (upper row) and of the bulk viscosity at
    $\eta_0=0.01$ (lower row). Lines
    are the least-squares fits, whose slopes are the damping sensitivities
    quoted in the main text.}
    \label{fig:compactness_rates}
\end{figure*}

We now turn to discuss the compactness sequence of Table~\ref{tab:compactness}. Figures~\ref{fig:compactness} and~\ref{fig:compactness_rates} show the results along this compactness sequence at $r_p/r_e=0.85$. Its model with $C=0.144$ coincides with the $r_p/r_e=0.85$ model of the main sequence, which we use as reference.

Both figures show that damping grows strongly with compactness. The amplitudes decay much faster in more compact stars (Fig.~\ref{fig:compactness}), and the dominant sensitivities increase by factors of 3 to 5 between the least and the most compact model (Fig.~\ref{fig:compactness_rates}). The division found along the rotation (main) sequence holds at every compactness: the $F$ mode is damped mainly by bulk viscosity, the $\fmode{2}$ and $\fmode{3}$ modes by shear viscosity, and the $\fmode{1}$ only weakly (as explained before due to the nature of the $^1f$ mode). Shear viscosity is found to become particularly important for the damping of the $F$ mode in more compact stars: its shear decay rate grows from values consistent with zero in the least compact models to the point where the bulk damping exceeds it by only a factor of about two. In addition, the $\fmode{1}$ mode acquires a weak shear damping as well.

Much of this growth follows from the viscosity prescription, given by Eq.~\eqref{eq:visc}. The kinematic viscosity grows with the central sound speed, and the radius decreases with compactness, so that the viscosity increases along the compactness sequence. The coupling of each mode to its dominant viscosity changes only moderately with compactness; the main change is the growing shear damping of the $F$ mode.

\section{Discussion}
\label{sec:discussion}

We have studied the effects of bulk and shear viscosity on the damping of axisymmetric oscillations of uniformly rotating relativistic stars. To do so, we have implemented the BDNK formulation of viscous hydrodynamics in a two-dimensional numerical code that evolves the equations of motion on the fixed
spacetime of a rotating star (Cowling approximation). We have built two different sequences of stellar models, using those reported in~\cite{Font:2001_Cowling} as guidance, and we have studied how physical viscous damping is affected by the degree of rotation and compactness.

The damping of the axisymmetric modes of a rotating star is determined by the kind of motion in each mode. For the four modes considered in this work, we have found that the fundamental quasi-radial $F$ mode is damped by bulk viscosity, as it corresponds to a mode that compresses the star globally. The $l=2$ and $l=3$ modes are nearly incompressible and are damped by shear viscosity. Finally, the $l=1$ mode, whose flow is close to a uniform translation, is barely damped at all by either shear and bulk viscosity. Rotation mixes these behaviours, most clearly by making the $F$ mode sensitive to shear. This qualitative picture, which is the main result of our study, does not depend on the details of the simulations. It holds for all rotation rates and degree of stellar compactness we have considered, at all resolutions tested, and in the extrapolation to infinite numerical resolution. We note, however, that the values of the damping rates and sensitivities reported in this work should be read as indicating the relative importance of shear and bulk viscosity for each mode, not as actual predictions of realistic damping times. Bulk viscosity from weak interactions out of equilibrium may be large enough to affect the dynamics of hot merger remnants~\cite{Alford:2018,Most:2022_bulk}. Our results, albeit obtained for uniformly rotating models with simplistic EoS, suggest that it would act mainly on the quasi-radial oscillations and much less on the quadrupolar $f$ mode that dominates the GW signal.

This study has several limitations. The Cowling approximation ignores the metric perturbations, which change the frequencies of the quasi-radial modes and remove the coupling to gravitational radiation. The viscosity prescription we have employed (cf.~Eq.~\eqref{eq:visc}) is a convenient model rather than a microphysical one; with realistic transport coefficients the damping rates will change, although the division between bulk-damped compressional modes and shear-damped incompressible modes might remain, as it does along our compactness sequence. On the numerical part, improving the consistency between primitive and conserved variables in the outer layers of the star, as long as testing more EoS and different rotation profiles are obvious next steps to investigate.

Our study has also shown that the inherent unphysical (numerical) damping affects only the constant term of the linear fits of the decay rates we have computed, not their slopes, and therefore does not contaminate the viscous sensitivities, which capture the physical damping. This is a crucial point of the present study, whose primary goal has been to characterize qualitatively the role of viscosity, as described by the BDNK framework, in the oscillations of rotating neutron stars. While this goal has been accomplished we note that the numerical values of the damping rates reported here should be regarded as upper limits at best, and are not the central result of our study. The key outcome of this work is a set of viscous sensitivities that reveal which type of viscosity, shear or bulk, dominates the damping of each oscillation mode, and establish robust trends as functions of rotation rate and compactness. These trends provide a qualitative map of viscous damping across parameter space that can serve as a foundation for more quantitative studies in the future.

\section*{Acknowledgments}
We thank Alex Pandya for his help during the early stages of the code’s development, and Harry L. H. Shum for useful discussions.
OHP acknowledges support by the Generalitat Valenciana and the European Social Fund (ESF) CIACIF 475/2024. This research is supported by the Spanish Agencia Estatal de Investigación (grant PID2024-159689NB-C21) funded by MICIU/AEI/10.13039/501100011033 and by FEDER / EU, by the Generalitat Valenciana (Prometeo Excellence Programme grant CIPROM/2022/49), and by the European Horizon Europe staff exchange (SE) programme HORIZON-MSCA-2021-SE-01 (Grant No.-NewFunFiCO-101086251). The simulations were performed on the MareNostrum supercomputer at the Barcelona Supercomputing Center (BSC), through allocations granted by the Red Española de Supercomputación (RES, AECT-2026-1-0040).

\appendix

\section{Resolution study}
\label{app:convergence}

\begin{figure}[t!]
    \centering
    \includegraphics[width=\linewidth]{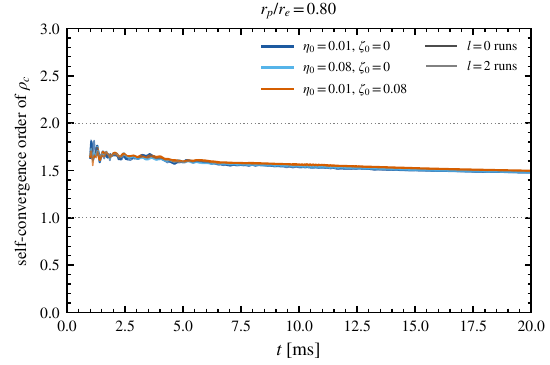}
    \caption{Time evolution of the self-convergence order of the central density $\rho_c$, Eq.~\eqref{eq:selfconv}, for the
    three viscous configurations of the $r_p/r_e=0.80$ model, with $l=0$ (thick) and $l=2$ (thin
    lines). The first millisecond is omitted. Dotted lines mark first and second order convergence.}
    \label{fig:convergence}
\end{figure}

For our resolution study we evolve the $r_p/r_e=0.80$ model with both $l=0$ and $l=2$ perturbations at three different resolutions, $N=144$, $216$ and $324$ cells. We consider perfect fluid evolutions and viscous runs, in the latter case with coefficients  $(\eta_0,\zeta_0)=(0.01,0)$, $(0.08,0)$ and $(0.01,0.08)$. All numerical settings (solver, cell reconstruction, primitive recovery, etc) are identical at all three resolutions. For global quantities we measure the self-convergence order $p$, which does not require the exact solution. If the error of a quantity $q$ is proportional to $h^p$, where $h\propto1/N$ is the grid spacing, the differences between successive resolutions give
\begin{equation}
    p=\frac{\ln\left(\left|q_{144}-q_{216}\right|/\left|q_{216}-q_{324}\right|\right)}{\ln 1.5},
    \label{eq:selfconv}
\end{equation}
since the resolution increases by the same factor, $1.5$, at each step. We evaluate Eq.~\eqref{eq:selfconv} at every time step.

The central density of the BDNK evolutions is the quantity most affected by resolution, because its error is dominated by the re-adjustment of the star described in Sec.~\ref{sec:transient}. As shown in Fig.~\ref{fig:convergence} its observed convergence order is about $1.5$, nearly constant in time. The same order is found for all viscosity coefficients considered  and for both multipoles. This value lies between the second order of the scheme in the interior of the star and the first order of the treatment of the surface, so that both contribute to the error at these resolutions.

The models with $r_p/r_e=1.00$ and $0.65$, which we also evolved at the three resolutions for the extrapolations of Sec.~\ref{sec:transient}, give consistent results: their mode frequencies change by less than $0.1\%$ between resolutions, and the ordering of shear and bulk damping is the same at all resolutions.

\bibliographystyle{apsrev4-2}
\bibliography{bib}

\end{document}